\documentclass[9pt,english,12pt, draftclsnofoot, onecolumn]{IEEEtran}
\usepackage[T1]{fontenc}
\usepackage[latin9]{inputenc}
\usepackage{amstext}
\usepackage{graphicx}

\makeatletter
\usepackage{balance}
\usepackage[nospace]{cite}
\usepackage{lipsum}

\makeatother

\usepackage{babel}
\begin{document}

\title{Analog MIMO Radio-over-Copper: Prototype and Preliminary Experimental
Results}

\author{A. Matera\thanks{A. Matera, M. Donati and U. Spagnolini are with DEIB, Politecnico
di Milano (E-mails: \{andrea.matera, umberto.spagnolini, marcello.donati\}@polimi.it).}, V. Rampa\thanks{V. Rampa is with Institute of Electronics, Computer and Telecommunication
Engineering (IEIIT), CNR, Italy (E-mail: vittorio-rampa@ieiit.cnr.it).}, M. Donati, A. Colamonico\thanks{A. Colamonico is with ABC Progetti Milano (E-mail: armando.colamonico@tin.it).},
A. F. Cattoni\thanks{A. Cattoni is with Keysight Technologies Denmark (E-mail: andrea.cattoni@keysight.com).}
and U. Spagnolini\thanks{This project has been partially funded by the TRIANGLE project, European
Union\textquoteright s Horizon 2020 research and innovation programme,
grant agreement No 688712.}}
\maketitle
\begin{abstract}
Analog Multiple-Input Multiple-Output Radio-over-Copper (A-MIMO-RoC)
is an effective all--analog FrontHaul (FH) architecture that exploits
any pre-existing Local Area Network (LAN) cabling infrastructure of
buildings to distribute Radio-Frequency (RF) signals indoors. A-MIMO-RoC,
by leveraging a fully analog implementation, completely avoids any
dedicated digital interface by using a transparent end-to-end system,
with consequent latency, bandwidth and cost benefits. Usually, LAN
cables are exploited mainly in the low-frequency spectrum portion,
mostly due to the moderate cable attenuation and crosstalk among twisted-pairs.
Unlike current systems based on LAN cables, the key feature of the
proposed platform is to exploit more efficiently the huge bandwidth
capability offered by LAN cables, that contain 4 twisted-pairs reaching
up to 500 MHz bandwidth/pair when the length is below 100 m. Several
works proposed numerical simulations that assert the feasibility of
employing LAN cables for indoor FH applications up to several hundreds
of MHz, but an A-MIMO-RoC experimental evaluation is still missing.
Here, we present some preliminary results obtained with an A-MIMO-RoC
prototype made by low-cost all-analog/all-passive devices along the
signal path. This setup demonstrates experimentally the feasibility
of the proposed analog relaying of MIMO RF signals over LAN cables
up to 400 MHz, thus enabling an efficient exploitation of the LAN
cables transport capabilities for 5G indoor applications. 
\end{abstract}


\thispagestyle{empty} 

\section{Introduction}

Cloud Radio Access Network (C-RAN) is a very attractive architecture
to handle a large number of users and antennas in the same radio resources,
as demanded by 5G mobile networks \cite{Boccardi-Heath-etal_2014}.
In particular, C-RAN is enabled by: \textit{i)} the colocation of
BaseBand Units (BBUs) in so-called BBU pools, where centralized processing
and interference mitigation are performed; \textit{ii)} the remote
placement of antennas and Radio-Frequency (RF) equipment (Remote Radio
Unit, RRU) that are moved away from the BBU and geographically distributed
both indoors and outdoors close to the end-users. This allows for
centralized signal processing, network scalability, increased spectral
efficiency and costs reduction \cite{Checko-Christiansen-etal_2015}.
C-RAN is already deployed in current (4G) mobile networks, in which
BBUs and RRUs exchange digital In-phase and Quadrature (I/Q) signals
over the so-called FrontHaul (FH) link according to any of the routinely
employed communication protocols e.g., the Common Public Radio Interface
(CPRI) \cite{CPRI:specs}. However, the massive number of antennas
and the increase in radio signal bandwidth expected for 5G networks
call into question the effectiveness of digital I/Q streaming, which
would introduce a $\text{\ensuremath{\approx}32}\times$ bandwidth
expansion over the FH link. A viable solution is to completely overtake
any signals' digitization at the RRUs in favor of a fully analog FH
where the RRUs up/down-convert and directly relay the analog RF signals
to/from the BBU. Analog FH is a promising solution for ultra-low latency
applications where latency is ultimately limited to signal propagation,
and also enables precise bit/carrier-frequency synchronization among
multiple RRUs for MIMO joint processing. \cite{Gambini-Spagnolini_2013,RadioDots2014,wake:RoFdesign,PWMglobecom}. 

In this context, C-RAN with analog FH based on Local Area Network
(LAN) cables, namely Analog Radio-over-Copper (A-RoC), has been shown
to be a cost-effective/bandwidth-efficient solution for upcoming 5G
indoor networks \cite{Naqvi-Matera-etal_2017,Tonini-Fiorani-etal_2017}.
\begin{figure}
\begin{centering}
\includegraphics[width=0.75\columnwidth]{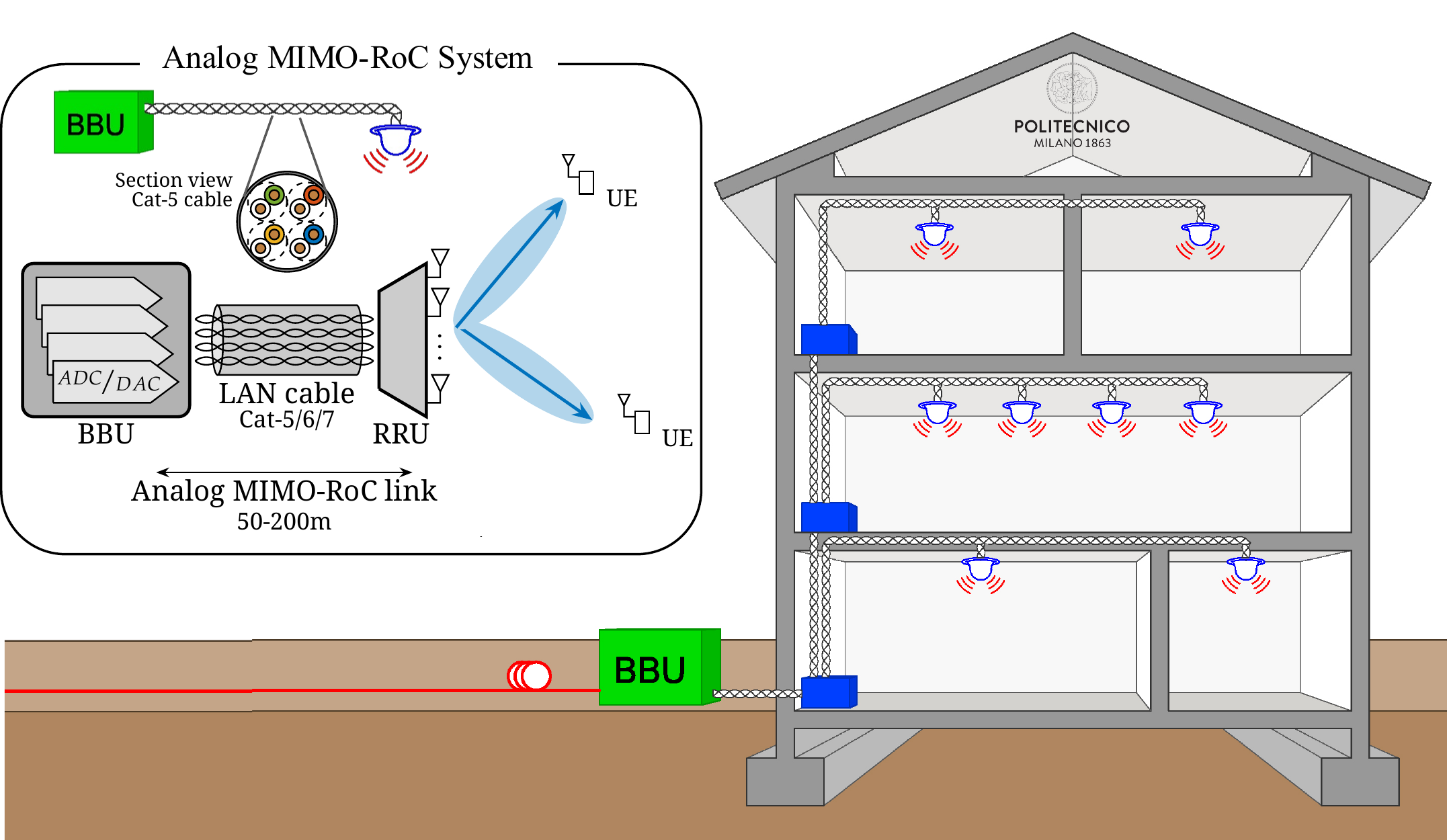}
\par\end{centering}
\caption{\label{fig:AMIMORoC architecture}Analog MIMO Radio-over-Copper architecture}
\end{figure}
 The A-RoC concept has been first proposed for femto-cell systems
to exchange analog RF signals between a remote location hosting all
PHY/MAC functionalities (BBU) and an in-home antenna device performing
only the analog relay of signals (RRU) \cite{Gambini-Spagnolini_2013}.
Afterwards, A-RoC gained lots of attention becoming the basis of commercial
solutions exploiting the pre-existing LAN cables of buildings to extend
indoor coverage over distances longer than 100 m \cite{RadioDots2014}.
By using all 4 twisted-pairs contained into the LAN cable at low-frequency
(characterized by low attenuation and crosstalk interference), one
can serve up to 4 antennas (e.g., 4x4 MIMO) per LAN cable. An A-RoC
architecture for LTE signals, referred to as LTE-over-Copper, has
been proposed in \cite{Medeiros-Huang-etal_2016} and references herein,
showing the capability to relay 3GPP-compliant LTE signals in the
21-24 MHz cable frequency band. 

Although demonstrating the feasibility of the A-RoC architecture,
the aforementioned works consider only the low-frequency portion of
copper-cables, i.e., $\approx$ 0-120 MHz, due to low cable attenuation
and crosstalk among twisted-pairs, thus not fully exploiting the transport
capability of LAN cables. In this direction, A-RoC has been recently
extended to the so-called A-MIMO-RoC architecture (see Fig. \ref{fig:AMIMORoC architecture}),
precisely with the goal to make a more efficient usage of the large
bandwidth offered by the 4 twisted-pairs contained into a single LAN
cable, providing up to 500 MHz bandwidth/pair (2 GHz bandwidth overall)
\cite{Naqvi-Matera-etal_2017,Matera-Combi-etal_2017,matera2017optimal,Matera-Spagnolini_2018,Matera-Spagnolini_2019,Rizzello-Joham-etal_2019}.
Simulation results demonstrated that, in principle, a single 100-m
Cat-6 LAN cable is capable to serve up to 60 RRU antennas each carrying
a 20-MHz LTE signal, achieving up to 10 Gbps equivalent wireless data-rate
\cite{Naqvi-Matera-etal_2017}. However, it has been demonstrated
that in order to achieve this performance figure, it is needed to
define an opportune low-complexity/all-analog mapping of the RF signals
at the RRU antennas onto a combination of twisted-pairs/frequency
channels over the cable. This wired-wireless resource allocation strategy,
denoted as Space-Frequency to Space-Frequency (SF2SF) multiplexing,
has been shown to minimize interference among signals carried over
different twisted-pairs, hence enabling an efficient usage of the
LAN cable bandwidth \cite{Matera-Spagnolini_2018,Matera-Combi-etal_2017,matera2017optimal,Matera-Spagnolini_2019}.
An alternative formulation/solution to the RoC wired-wireless resource
scheduling problem has been recently presented for 4G systems \cite{Naqvi-Jabeen-Ho_2018},
and then extended to 5G New Radio (NR) indoor architectures \cite{Naqvi-Ho-Peng_2018,Naqvi-Ho-Shahida_2018}. 

Previous works \cite{Naqvi-Matera-etal_2017,Matera-Combi-etal_2017,matera2017optimal,Matera-Spagnolini_2018,Matera-Spagnolini_2019,Rizzello-Joham-etal_2019}
proved the effectiveness of A-MIMO-RoC in terms of equivalent wireless
capacity by numerical analysis, however, an experimental validation
of the A-MIMO-RoC architecture is still missing, and this is the focus
of this paper. It is worth mentioning that the proposed A-MIMO-RoC
platform is also suitable in antenna remotization scenarios for indoor
WLAN applications where the cost of CPRI interfaces cannot be afforded.

\subsubsection*{Contribution}

This paper presents preliminary experimental results demonstrating
the potential of A-MIMO-RoC as an effective architecture for perspective
5G indoor networks. A prototype has been purposely developed in order
to evaluate the performance of A-MIMO-RoC whereby multiple LTE signals
are carried over the same LAN cable at high cable frequency, i.e.,
impaired by severe intra-cable crosstalk and cable attenuation \cite{Naqvi-Matera-etal_2017}.
In particular, experiments demonstrate the feasibility of transporting,
in an all-analog fashion, multiple MIMO LTE signals over a single
multi-pair copper-cable up to 400 MHz cable frequency with negligible
performance degradation. This enables a more efficient exploitation
of LAN cables transport capabilities, which are thus proved to offer
enough bandwidth for indoor applications. To this aim, the TRIANGLE
testbed \cite{cattoni2016end,triangle_websit} played a key role in
measuring the end-to-end performance degradation introduced by the
proposed A-MIMO-RoC architecture. 

\subsubsection*{Organization}

The paper is organized as follows: Sect. \ref{sec:The-Triangle-Testbed}
introduces the TRIANGLE testbed used for the experiments, while the
setup for testing the A-MIMO-RoC platform is in Sect. \ref{sec:Dual-RoC-Experiment-Setup}.
Sect. \ref{sec:Experimental-Results} shows some preliminary experimental
results. Sect. \ref{sec:Conluding-Remarks-=000026} draws some preliminary
conclusions and highlights future research directions.

\section{\label{sec:The-Triangle-Testbed}The TRIANGLE Testbed}

\begin{figure*}
\begin{centering}
\includegraphics[width=0.75\textwidth]{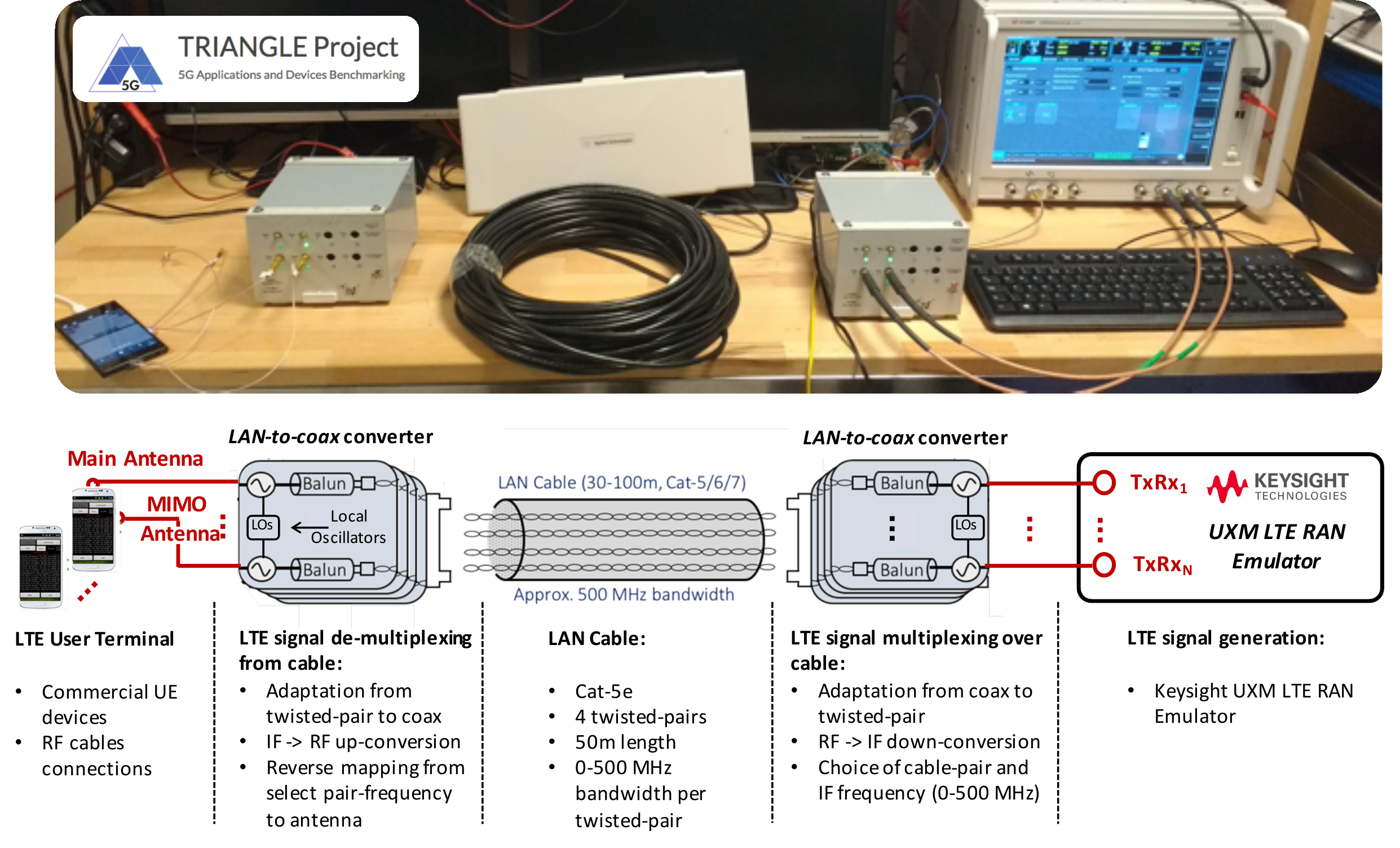}
\par\end{centering}
\caption{\label{fig:LTE-alone-Experiment-1}Experimental Setup for A-MIMO-RoC}
\end{figure*}
The TRIANGLE testbed is the main core of the H2020 TRIANGLE project
\cite{triangle_websit}, whose objective is to promote the testing
and benchmarking of 5G mobile applications and User Equipments (UEs).
In particular, the TRIANGLE testbed allows to run multiple tests in
a controlled environment that encompasses all elements of the telecommunication
chain, from radio signal generation to the end-to-end testing of mobile
applications. An extensive description of the TRIANGLE testbed is
outside the scope of this paper. Details can be found in \cite{cattoni2016end,triangle_websit},
while here we review only the main elements that have been employed
for the A-MIMO-RoC testing.

\subsection{Keysight UXM RAN Emulator}

The UXM Wireless Test Platform by Keysight Technologies allows to
emulate multiple cellular networks in a controlled manner by supporting
multiple Radio Access Technologies (multi-RAT), including GSM/GPRS,
UMTS and LTE-Advanced networks (i.e., 2G, 3G, 4G and 4.5G). UXM features
include intra- and inter-RAT handovers, protocol debugging, IP end-to-end
delay, BLock Error Rate (BLER) and throughput measurements, in addition
to the possibility of performing RF conformance tests. The UXM device
offers great customization capabilities and an intuitive user interface,
which allow to easily evaluate the Key Performance Indicators (KPIs)
of interest under any possible system setting defined by the 3GPP.
Moreover, the UXM device features an advanced fading engine with the
main channels models defined by 3GPP \cite{3gpp_radio_tx}. The UXM
device has been employed to generate the MIMO LTE radio signals and
to emulate specific propagating channel conditions, e.g., the ETSI
Pedestrian at 5 Hz (EPA5) channel model, which fairly describes the
typical indoor office scenario targeted by the A-MIMO-RoC system. 

\subsection{Mobile Device Monitoring }

In order to measure the KPIs perceived by the end-device, some additional
tools, detailed below, are needed on the smartphones under test, i.e.,
the UE, according to the scenario in Fig. \ref{fig:LTE-alone-Experiment-1}.

\subsubsection{Performance Tool }

In order to guarantee high resolution reporting of the target QoS
KPIs, especially in 5G scenarios, the DEKRA Performance Tool has been
integrated into the TRIANGLE testbed \cite{DEKRA}. Beside providing
accurate one-way measurements, the DEKRA Performance Tool tool includes
a built-in traffic generator and enables the automation and testing
of Android mobile Apps by measuring relevant Quality of Experience
(QoE) KPI. In particular, the DEKRA Performance Tool was employed
to perform end-to-end YouTube application testing, which allowed us
to quantify the performance degradation introduced by the relaying
over LAN cable in terms of YouTube video streaming quality. However,
these results have not been reported here for brevity.

\subsubsection{TestelDroid Mobile Monitoring App }

TestelDroid is a software tool developed by the University of Malaga
(UMA) \cite{TestelDroid} that enables passive monitoring of radio
parameters and data traffic in Android-based devices. Logging is implemented
as an Android service that can be running in the background logging
all information while running the application under test, e.g., YouTube.
This functionality enables monitoring of traffic information generated
by any application, which extends the testing to a very wide range
of use cases. The logged parameters (neighbor cells, GPS, traffic,
etc.) can be flexibly configured. 

\section{\label{sec:Dual-RoC-Experiment-Setup}A-MIMO-RoC Experiment Setup }

To test the A-MIMO-RoC platform, we used the TRIANGLE testbed in the
typical device-testing configuration \cite{triangle_websit}, but
inserting a 4-pairs RJ45 Cat-5e LAN cable between the RF output ports
of the UXM and the RF connections at the UE (see Fig. \ref{fig:LTE-alone-Experiment-1})
to evaluate experimentally the performance degradation due to the
all-analog RF-to-RF relaying of signals over copper. As shown in Fig.
\ref{fig:LTE-alone-Experiment-1}, the RF signals from the UXM are
transported by two LAN-to-Coax Converters (L2CCs) to the remote antennas
and then to the UEs. For lack of space, a detailed circuit description
of the two converters, that represent the core of the A-MIMO-RoC platform,
is omitted here. L2CCs are two identical devices connected back-to-back
through the RJ45 LAN cable. The L2CC on the right of Fig. \ref{fig:LTE-alone-Experiment-1}
performs impedance adaptation, RF down-conversion, Intermediate-Frequency
(IF) mapping and IF cable equalization to convert the RF signals into
their IF counterparts to feed the 4 twisted-pairs cables. The L2CC
on the left performs the inverse operations on the IF signals: cable
equalization, IF demapping, RF up-conversion and impedance adaptation
to recover the RF signals. The passive cable equalization circuit,
designed for 50-m and 100-m LAN cables, is split between both L2CCs,
while adaptive active equalization is envisioned for a future version
of the prototype. Both L2CCs include only low-cost/complexity all-passive
all-analog devices along the signal path in order to perform fully
bi-directional transparent operations, e.g., for Time Division Duplexing
(TDD) systems. For all wireless communication trials with the A-MIMO-RoC
platform, we used a 50-m long Cat-5e LAN cable, commonly deployed
in buildings, with a cable bandwidth tested here up to 400 MHz/pair.

The experimental setup is represented in Fig. \ref{fig:LTE-alone-Experiment-1},
together with a simplified block diagram detailing the role of each
of the components used for the experiments (i.e., RAN Emulator, L2CCs,
LAN cables and UE). In details, the experiment setup is as follows
(only the downlink is described, uplink symmetrical): \textit{(i)}
up to 4 RF LTE signals are generated by the UXM; \textit{(ii)} RF
cables are connected at each RF output ports of the UXM; \textit{(iii)}
the signal carried on each RF cable is down-converted to IF to match
the bandwidth over the LAN cable, e.g., in the 10-400 MHz frequency
range, possibly multiplexed in frequency over cable by the first L2CC;
\textit{(iv)} each IF-converted signal is conveyed by one of the 4
twisted-pairs: cable adaptation/equalization, coax-to-pairs mapping
and RF/IF down-conversion between coax and twisted-pair is performed
by the L2CC; \textit{(v)} at the other end of the LAN cable, the second
L2CC performs cable adaptation/equalization, pairs-to-coax de-mapping
and IF/RF up-conversion to interface with the UE under test by using
RF cables; \textit{(vi)} DEKRA Performance Tool and TestelDroid, integrated
into the TRIANGLE testbed, are used to test the UEs.

\section{\label{sec:Experimental-Results}Experimental Results}

The focus of this experiment was to demonstrate the viability of remotizing
RF antennas by relaying multiple LTE signals over the same 50-m Cat-5e
LAN cable and to evaluate the end-to-end performance degradation introduced
by the cable. It is well known that the low-frequency spectrum of
LAN cables is suitable for transmission due to the mild crosstalk
among pairs and cable attenuation (see \cite{Naqvi-Matera-etal_2017}
for typical LAN cables characteristics). However, in order to enable
a more efficient usage of the LAN cable bandwidth, our goal is to
prove that also the more challenging high-frequency spectrum of LAN
cables can be exploited for the analog FH transmission. 

\begin{table}
\vspace{-10pt}

\caption{\label{tab:LTE-system-settings}LTE system settings}

\begin{centering}
\includegraphics[width=0.5\columnwidth]{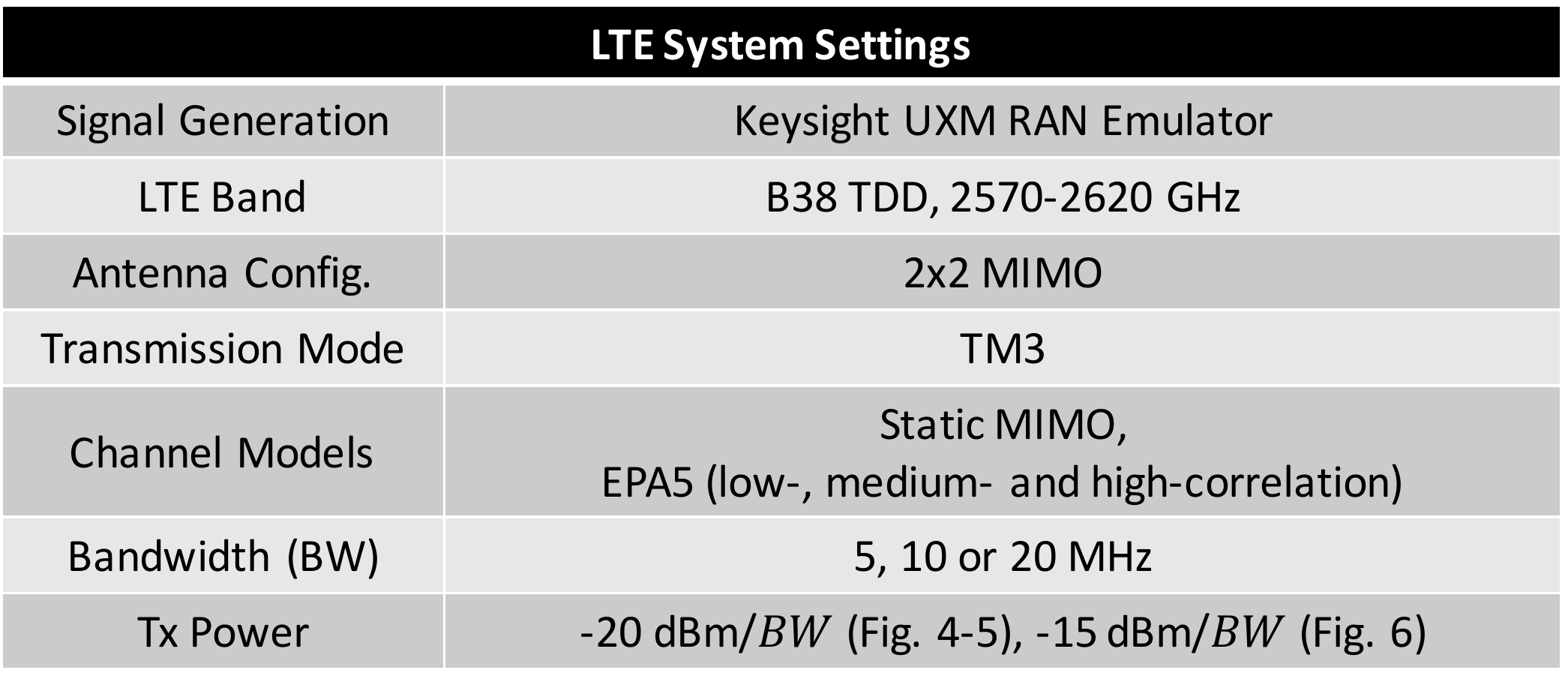}
\par\end{centering}
\vspace{-25pt}
\end{table}
 In particular, we tested the relaying over copper-cable of a $2\times2$
MIMO LTE signal using the LTE system settings parameters reported
in Table \ref{tab:LTE-system-settings}. Performance have been evaluated
in the DownLink (DL) direction in terms of throughput and BLER for
different Modulation and Coding Schemes (MCS) values, ranging from
0 (0-QPSK) to 17 (17-16QAM) \cite{3gpp_PHY_procedure}. In all experiments,
we run the tests by keeping fixed each MCS configuration for a duration
of $\approx2$ minutes. Each measurement point of plots is the average
of the given performance metric over 500 ms of test to obtain approx.
240 measurements points for each MCS. Two main experiments, detailed
in the following, have been performed for testing the A-MIMO-RoC platform.

\subsection{A-MIMO-RoC and cable IF }

\begin{figure}
\begin{centering}
\includegraphics[width=0.65\columnwidth]{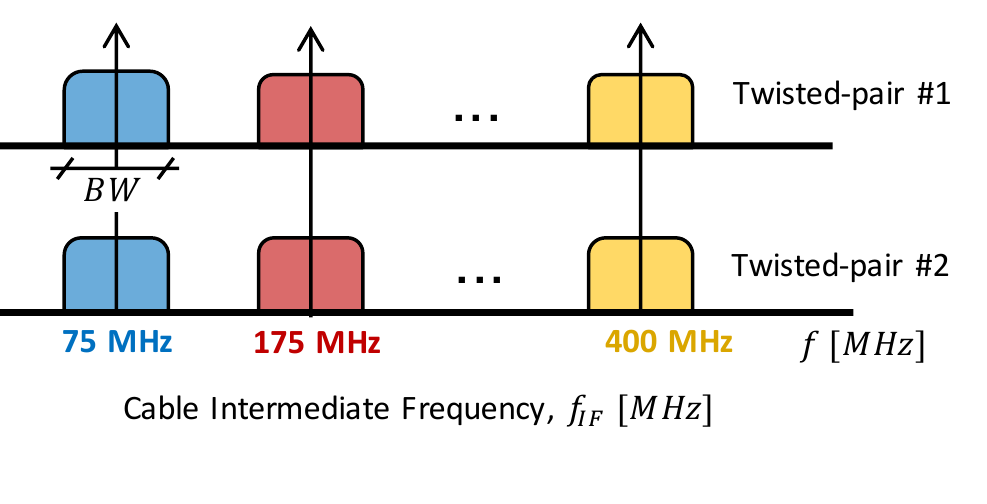}
\par\end{centering}
\caption{\label{fig:LTE-signal-mapping}RF signal mapping over cable: 2 LTE
bands with bandwidth $BW$ mapped over two different cable-pairs at
the same cable-IF.}
\end{figure}

The goal of this experiment was two-fold: \textit{i)} to prove the
feasibility of transporting the 2 RF bands, corresponding to the 2
MIMO ports, over 2 different twisted-pairs of the LAN cable but at
the same cable IF frequency $f_{\text{IF}}$, where each pair interferes
with the other one (see Fig. \ref{fig:LTE-signal-mapping}); and \textit{ii)}
to evaluate the performance degradation caused by increasing the value
$f_{\text{IF}}$, i.e., to increase the interference among pairs and
the attenuation levels \cite{Naqvi-Matera-etal_2017}. 

\begin{figure}
\begin{centering}
\includegraphics[width=0.75\columnwidth]{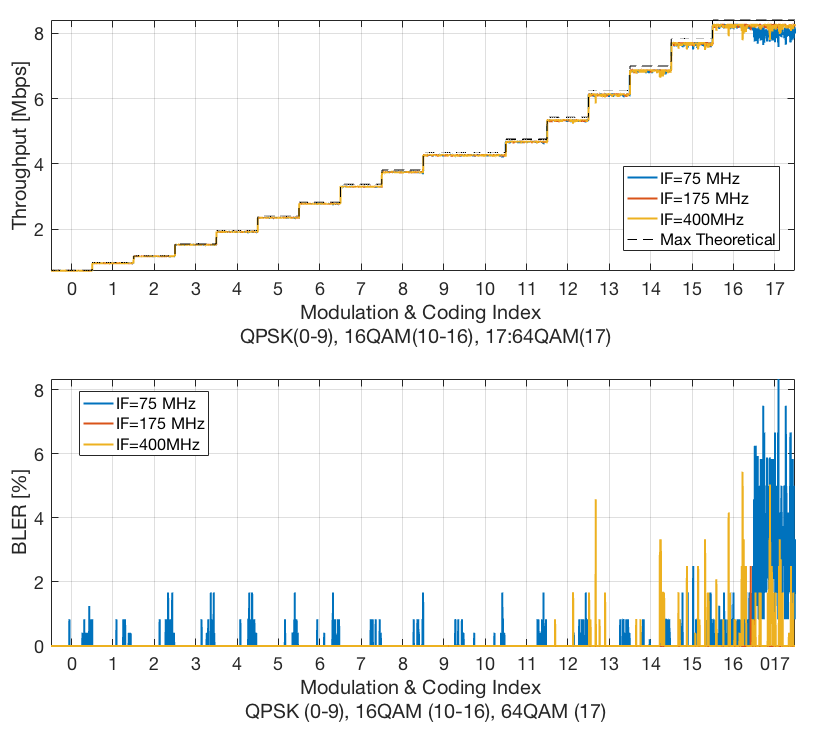}
\par\end{centering}
\caption{\label{fig:MIMO-LTE-over-Copper-performance}A-MIMO-RoC performances
(DL throughput and BLER) vs. $f_{\text{IF}}$ for $BW=5$ MHz and
different modulation schemes. }
\end{figure}
\begin{figure}
\begin{centering}
\includegraphics[width=0.75\columnwidth]{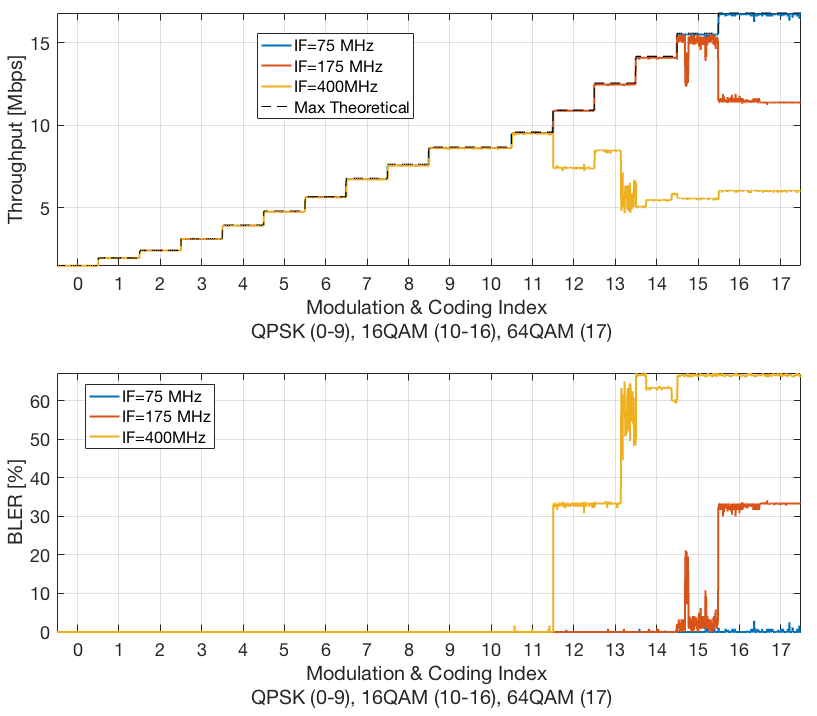}
\par\end{centering}
\caption{\label{fig:MIMO-LTE-over-Copper-performance-1}A-MIMO-RoC performances
(DL throughput and BLER) vs. $f_{\text{IF}}$ for $BW=10$ MHz and
different modulation schemes. }
\end{figure}
In Fig. \ref{fig:MIMO-LTE-over-Copper-performance}-\ref{fig:MIMO-LTE-over-Copper-performance-1},
throughput and BLER are shown considering a Static MIMO channel at
-20 dBm/$BW$ input power by using three different cable IF values,
as shown in Fig. \ref{fig:LTE-signal-mapping}: $f_{\text{IF}}=75$
MHz (blue curves), $f_{\text{IF}}=175$ MHz (red curves) and $f_{\text{IF}}=400$
MHz (yellow curves). The maximum theoretical throughput achievable
by each MCS \cite{3gpp_PHY_procedure} (black dashed line) over the
considered channel bandwidth ($BW$) is shown as reference, thus quantifying
the performance degradation introduced by the proposed system. For
the $BW=5$ MHz case, shown in Fig. \ref{fig:MIMO-LTE-over-Copper-performance},
the loss in terms of throughput introduced by the A-MIMO-RoC setup
is almost negligible for all considered MCSs and $f_{\text{IF}}$
values. As expected, BLER increases for high MCS, but the degradation
with respect to the maximum achievable throughput is still small.
However, for $BW=10$ MHz shown in Fig. \ref{fig:MIMO-LTE-over-Copper-performance-1},
the maximum throughput is achieved for all MCSs only for $f_{\text{IF}}=75$
MHz, while for $f_{\text{IF}}=175$ MHz and $f_{\text{IF}}=400$ MHz
the maximum MCS values that can be employed with fairly low BLER are
15-16QAM and 11-16QAM, respectively. This performance degradation,
which is more pronounced for higher $f_{\text{IF}}$, is totally expected,
and it is explained by the fact that both cable crosstalk and attenuation
levels increase with cable frequency \cite{Naqvi-Matera-etal_2017}.
Moreover, the transmit power is the same for both settings, i.e.,
$BW=5$ MHz and $BW=10$ MHz (see Table \ref{tab:LTE-system-settings}).
Hence, the smaller channel bandwidth, the higher Signal-to-Noise Ratio
(SNR) per subcarrier, which explains the better performance obtained
in the $BW=5$ MHz case. Notice that, as both L2CCs are all-passive
and bi-directional devices, signal relaying over cable introduces
a significant attenuation (in the order of 60 dB for both L2CCs) which
forces the whole system to work close to the UE sensitivity \cite{3gpp_PHY_procedure},
even by setting the transmit power to the maximum value allowed by
the hardware devices. It is thus expected that the performance loss
observed in case of $BW=10$ MHz can be avoided by introducing some
signal amplification in the system design, which is left as future
works. In any case, Fig. \ref{fig:MIMO-LTE-over-Copper-performance-1}
confirms the feasibility of relaying LTE signals over high-frequency
copper-cable bands: even in the worst case of 2 LTE bands carried
over 2 different twisted-pairs at the same $f_{\text{IF}}=400$ MHz,
it is still possible to achieve a throughput of $\approx10$ Mbps
over $BW=10$ MHz. Of course, WLAN indoor scenarios with shorter cables
(e.g., length $\approx$ 10-20 m) can fully exploit the transparent
antenna remotization capabilities and performances of the A-MIMO-RoC
platform.

\subsection{A-MIMO-RoC and Channel Models }

The goal of this experiment was to evaluate the performance of the
A-MIMO-RoC platform for different MIMO channel models. In particular,
we adopted the EPA5 channel \cite{3gpp_radio_tx}, as it fairly describes
the propagating conditions of the indoor office environment targeted
by the A-MIMO-RoC platform. We considered a $2\times2$ MIMO LTE signal
with $BW=20$ MHz, where the 2 RF bands were transported over 2 different
twisted-pairs at the same frequency $f_{\text{IF}}=175$ MHz. In this
experiment, the input power was set to -15 dBm/$BW$.

\begin{figure}
\begin{centering}
\includegraphics[width=0.75\columnwidth]{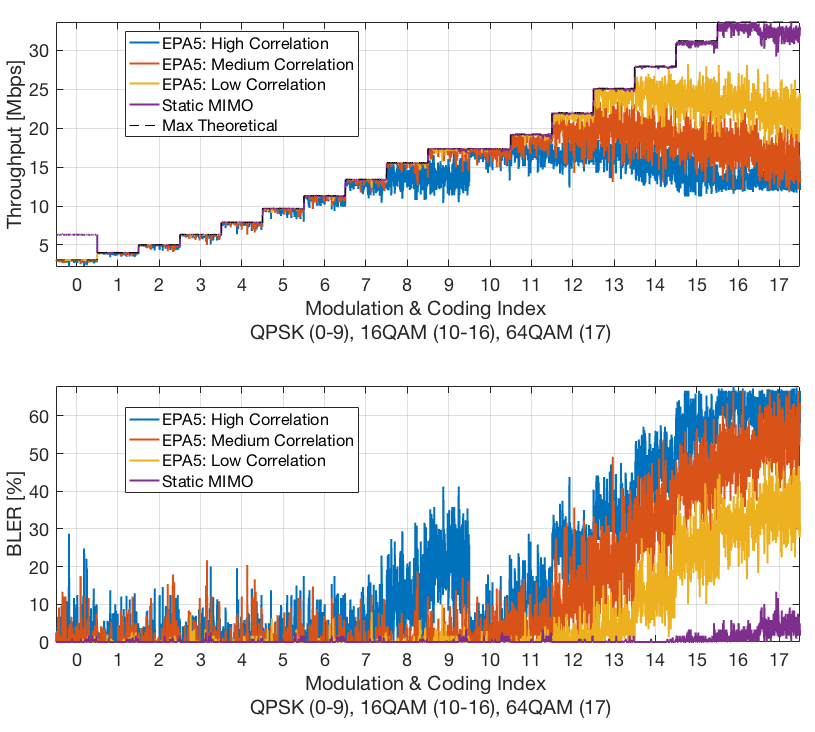}
\par\end{centering}
\caption{\label{fig:MIMO-LTE-over-Copper-vs}A-MIMO-RoC vs Channel Model: DL
throughput and BLER versus and MCS, $BW=20$ MHz}
\end{figure}
 Fig. \ref{fig:MIMO-LTE-over-Copper-vs} shows that, in the case of
Static MIMO channel, even for $BW=20$ MHz, the performance degradation
introduced by the cable is negligible (apart from a small loss for
high MCS). However, for the EPA5 channel model, performance get worse,
especially for medium and high channel correlation and for high MCS.
In fact, for practical applications, BLER should be below 10-15\%:
higher BLER would imply continuous HARQ/ARQ retransmissions leading
to unacceptable end-to-end latency, and thus nullifying the latency
benefits of the all-analog FH. Accordingly, Fig. \ref{fig:MIMO-LTE-over-Copper-vs}
shows that for low- and medium-correlated channels, performance for
practical applications are limited to 16QAM with $\approx17$ Mbps
throughput, while they are further reduced to QPSK ($<15$ Mbps) for
high-correlated MIMO channel. Once again, the reason for this loss
relies in the all-passive circuitry implementation of our system,
that strongly simplifies the hardware design of the prototype but
introduces a significant attenuation. The power fluctuations introduced
by the EPA5 channel make the system oscillate around the UE sensitivity
threshold for high MCS, which explains the curves' behavior in the
rightmost portion of the figure. 

\section{\label{sec:Conluding-Remarks-=000026}Concluding Remarks \& Future
Works}

This paper presents some preliminary experimental results asserting
the feasibility of the Analog MIMO Radio-over-Copper (A-MIMO-RoC)
architecture to transparently relay multiple LTE signals over a single
LAN cable in an all-analog fashion, even at high cable frequency.
To this aim, an A-MIMO-RoC prototype platform with transparent RF-to-RF
capabilities has been purposely developed. The conclusions drawn by
the experiments can be summarized as follows: \textsl{i)} LAN cables
bandwidth capability can be exploited up to several hundreds of MHz
for transparently transporting MIMO Radio-Frequency (RF) signals indoors,
even when such signals are transported over different twisted-pairs
but at the same Intermediate-Frequency (IF), thus affected by strong
cable-crosstalk interference;\textit{ ii)} the performance degradation
experienced for high Modulation and Coding Schemes (MCSs) and high
IF values is mainly due to the low signal power received at the user
device caused by the attenuation introduced by the analog relay over
cable; \textit{iii)} again, due to the overall system attenuation,
the prototype developed appears to be quite sensitive to signal power
variations, and this reflects in performance loss when employing non-static
MIMO channel models, e.g., ETSI Pedestrian at 5 Hz. However, with
shorter LAN cable length e.g., for WLAN indoor applications, the overall
system attenuation problem is no more an issue. Concluding, although
the fully passive implementation considered here substantially simplifies
the hardware design, experiment results suggest that most of the issues
encountered might be solved by introducing some active circuitry such
as active equalization, amplification and power adaptations in the
LAN-to-coax converters. However, this is far from being trivial, and
it is left for future developments of our technology. Finally, further
experiments need to be conducted in order to evaluate other critical
metrics such as the end-to-end latency or more detailed packets' statistics.

\section*{Acknowledgement}

This project has been partially funded by the TRIANGLE project, European
Union Horizon 2020 research and innovation programme, grant agreement
No 688712.

\bibliographystyle{ieeetr}
\bibliography{Bibliography}

\begin{thebibliography}{10}

\bibitem{Boccardi-Heath-etal_2014}
F.~Boccardi {\em et~al.}, ``Five disruptive technology directions for {5G},''
  {\em IEEE Commun. Mag.}, vol.~52, no.~2, pp.~74--80, 2014.

\bibitem{Checko-Christiansen-etal_2015}
A.~Checko {\em et~al.}, ``Cloud {RAN} for mobile networks?{A} technology
  overview,'' {\em IEEE Commun. Surveys Tuts.}, vol.~17, no.~1, pp.~405--426,
  2015.

\bibitem{CPRI:specs}
``{{CPRI Specifications V.6.1 (2014-07-01)}},'' September 2014.

\bibitem{Gambini-Spagnolini_2013}
J.~Gambini {\em et~al.}, ``Wireless over cable for femtocell systems,'' {\em
  IEEE Commun. Mag.}, vol.~51, pp.~178--185, May 2013.

\bibitem{RadioDots2014}
C.~Lu {\em et~al.}, ``Connecting the dots: small cells shape up for
  high-performance indoor radio,'' {\em Ericsson Review}, vol.~91, no.~2,
  pp.~38--45, 2014.

\bibitem{wake:RoFdesign}
D.~Wake {\em et~al.}, ``Radio over fiber link design for next generation
  wireless systems,'' {\em J. Lightw. Technol.}, vol.~28, no.~16,
  pp.~2456--2464, 2010.

\bibitem{PWMglobecom}
L.~Combi {\em et~al.}, ``{{Pulse-Width optical modulation for CRAN
  front-hauling}},'' in {\em IEEE GLOBECOM}, (San Diego), Dec. 2015.

\bibitem{Naqvi-Matera-etal_2017}
S.~H.~R. Naqvi {\em et~al.}, ``{On the transport capability of LAN cables in
  all-analog MIMO-RoC fronthaul},'' in {\em Wireless Communications and
  Networking Conference (WCNC), 2017 IEEE}, pp.~1--6, IEEE, 2017.

\bibitem{Tonini-Fiorani-etal_2017}
F.~Tonini {\em et~al.}, ``Radio and transport planning of centralized radio
  architectures in {5G} indoor scenarios,'' {\em IEEE J. Sel. Areas Commun.},
  vol.~35, no.~8, pp.~1837--1848, 2017.

\bibitem{Medeiros-Huang-etal_2016}
E.~Medeiros {\em et~al.}, ``{{Crosstalk Mitigation for LTE-Over-Copper in
  Downlink Direction}},'' {\em IEEE Comm. Lett.}, vol.~20, pp.~1425--1428, July
  2016.

\bibitem{Matera-Combi-etal_2017}
A.~Matera {\em et~al.}, ``{Space-frequency to space-frequency for MIMO radio
  over copper},'' in {\em Communications (ICC), 2017 IEEE International
  Conference on}, pp.~1--6, IEEE, 2017.

\bibitem{matera2017optimal}
A.~Matera {\em et~al.}, ``On the optimal space-frequency to frequency mapping
  in indoor single-pair {RoC} fronthaul,'' in {\em Networks and Communications
  (EuCNC), 2017 European Conference on}, pp.~1--5, IEEE, 2017.

\bibitem{Matera-Spagnolini_2018}
A.~Matera {\em et~al.}, ``Analog {MIMO-RoC} downlink with {SF2SF},'' {\em IEEE
  Wireless Commun. Lett.}, pp.~1--1, 2018.

\bibitem{Matera-Spagnolini_2019}
A.~{Matera} {\em et~al.}, ``Analog {MIMO} radio-over-copper downlink with
  space-frequency to space-frequency multiplexing for multi-user {5G} indoor
  deployments,'' {\em IEEE Trans. Wireless Commun.}, vol.~18, pp.~2813--2827,
  May 2019.

\bibitem{Rizzello-Joham-etal_2019}
V.~{Rizzello} {\em et~al.}, ``Precoding design for the {MIMO-RoC} downlink,''
  in {\em ICASSP 2019 - 2019 IEEE International Conference on Acoustics, Speech
  and Signal Processing (ICASSP)}, pp.~4694--4698, May 2019.

\bibitem{Naqvi-Jabeen-Ho_2018}
S.~H. {Raza Naqvi} {\em et~al.}, ``Low-complexity optimal scheduler for {LTE}
  over {LAN} cable,'' in {\em 2018 International Conference on Networking and
  Network Applications (NaNA)}, pp.~36--41, Oct 2018.

\bibitem{Naqvi-Ho-Peng_2018}
S.~H. {Raza Naqvi} {\em et~al.}, ``A novel {5G} indoor service provisioning
  architecture,'' in {\em 2018 15th International Symposium on Wireless
  Communication Systems (ISWCS)}, pp.~1--6, Aug 2018.

\bibitem{Naqvi-Ho-Shahida_2018}
S.~H.~R. Naqvi {\em et~al.}, ``A novel distributed antenna access architecture
  for {5G} indoor service provisioning,'' {\em IEEE J. Sel. Areas Commun.},
  vol.~36, no.~11, pp.~2518--2527, 2018.

\bibitem{cattoni2016end}
A.~F. Cattoni {\em et~al.}, ``{An end-to-end testing ecosystem for 5G},'' in
  {\em Networks and Communications (EuCNC), 2016 European Conference on},
  pp.~307--312, IEEE, 2016.

\bibitem{triangle_websit}
{Triangle project official website}, ``www.triangle-project.eu.''

\bibitem{3gpp_radio_tx}
{3GPP TS 36.101 Group Radio Access Network; Evolved Universal Terrestrial Radio
  Access (E-UTRA)}, ``{User Equipment (UE) Radio Transmission and Reception}.''

\bibitem{DEKRA}
{DEKRA Performance Tool Official Website},
  ``https://wireless.dekra-product-safety.com/it-services-solutions.html.''

\bibitem{TestelDroid}
A.~Alvarez {\em et~al.}, ``Field measurements of mobile services with android
  smartphones,'' in {\em Consumer Communications and Networking Conference
  (CCNC), 2012 IEEE}, pp.~105--109, IEEE, 2012.

\bibitem{3gpp_PHY_procedure}
{3GPP TS 36.213 Group Radio Access Network; Evolved Universal Terrestrial Radio
  Access (E-UTRA)}, ``{Physical Layer Procedure}.''

\end{thebibliography}

\end{document}